\documentstyle[twocolumn,epsf]{jpsj}
\def\vS{{\mib S}}
\def\vT{{\mib T}}
\def\runtitle{Magnetization Plateau of an $S=1$ Frustrated Spin Ladder }

\title{Magnetization Plateau of an $S=1$ Frustrated Spin Ladder
}
\author{ Kiyomi {\sc Okamoto}$^1$, Nobuhisa {\sc Okazaki}$^2$ and
T\^oru {\sc Sakai}$^2$}
\inst{$^1$Department of Physics, Tokyo Institute of Technology, Meguro-ku, 
Tokyo 152-8551, Japan \\
$^2$Faculty of Science, Himeji Institute of Technology, \\
Kamigouri-cho, Akou-gun, Hyogo 678-1297, Japan \\ 
}
\recdate{~~~~~~~~~~~~~~~~~~~~~~~~~~~~~~}
\abst{
We study the magnetization plateau at $1/4$
of the saturation magnetization of the $S=1$ antiferromagnetic
spin ladder both analytically and numerically, 
with the aim of explaining recent experimental 
results on BIP-TENO by Goto {\it et al}.
We propose two mechanisms for the plateau formation and clarify the 
plateau phase diagram on the plane of the coupling constants between spins.
}

\kword{spin ladder, magnetization plateau, frustration}
\begin{document}
\sloppy
\maketitle

The magnetization plateaux of low dimensional spin systems
have been attracting much attention because of its full quantum nature.
Very recently, a new organic tetraradical, 
3,3',5,5'-tetrakis({\it N-tert}-butylaminoxyl)biphenyl, 
abbreviated as BIP-TENO, has been synthesized\cite{Katoh}.
This material can be regarded as an $S=1$ antiferromagnetic two-leg spin ladder,
as explained later.
Goto {\it et al.}\cite{Goto} measured the magnetization curve of BIP-TENO 
at low temperatures in pulsed high magnetic fields up to about 50 T.
They found $M \simeq 0$ up to about $H_{\rm c1} \simeq 10\,{\rm T}$,
suggesting the non-magnetic ground state,
which was consistent with the result of susceptibility measurement.
Another remarkable nature of their magnetization curve is the
plateau at $M=M_{\rm s}/4$ above $H_{\rm c2} \simeq 45\,{\rm T}$,
where $M_{\rm s}$ is the saturation magnetization.
Unfortunately the end of the $M_{\rm s}/4$ plateau is unknown
because of the limitation of the strength of their magnetic field.

Considering the structure of BIP-TENO,
Katoh {\it et al.}\cite{Katoh} proposed its model of Fig.1.
Since the ferromagnetic coupling $J_{\rm F}$ is thought to be much larger
than other antiferromagnetic couplings $J'_{\rm AF}$ and $J_{\rm AF1}$,
two spins coupled by $J_{\rm F}$ effectively form an $S=1$ spin.
Thus the model of Fig.1(b) is reduced to an $S=1$ antiferromagnetic two-leg
ladder, as shown in Fig.1(c).
The relation between the coupling constants of these two models are
\begin{equation}
    J_\perp = J_{\rm AF1},~~~~~
    J_1 = (1/4) J'_{\rm AF}.
\end{equation}
In the following, we consider the model of Fig.1(c)
to investigate the $M_{\rm s}/4$ magnetization plateau
of this model at zero temperature. 
\begin{figure}
\begin{center}
\leavevmode
\epsfxsize=5.0cm
\epsfbox{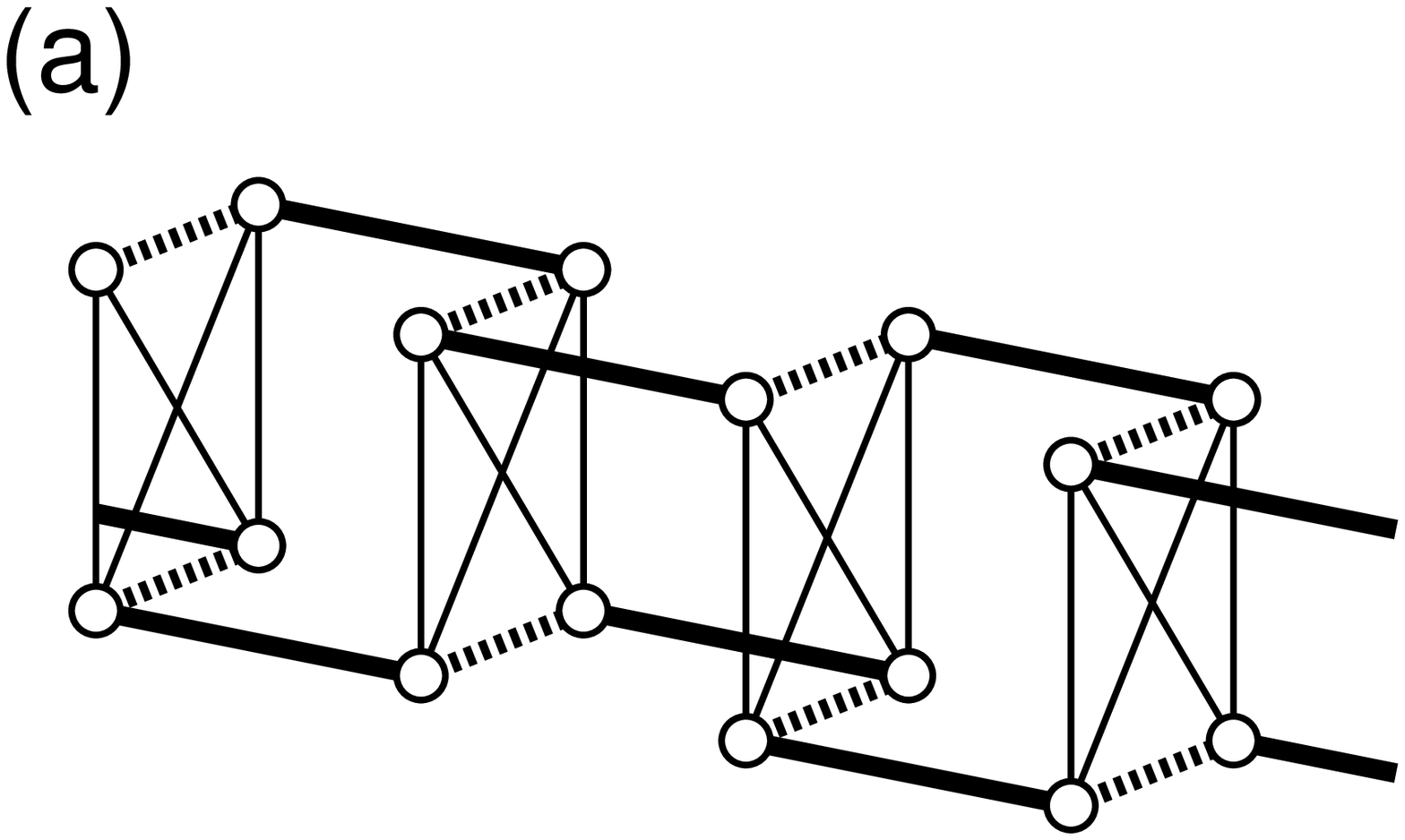}
\end{center}
\vskip0.3truecm
\begin{center}
\leavevmode
\epsfxsize=5.0cm
\epsfbox{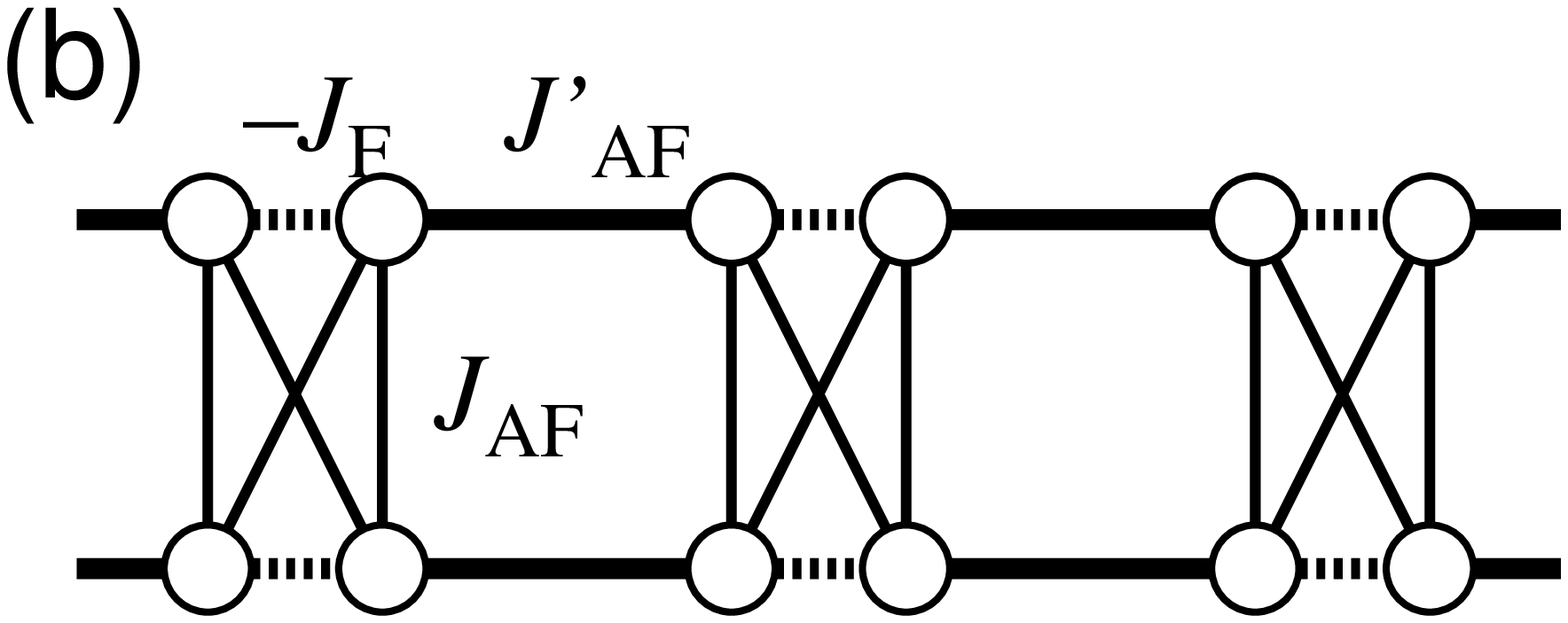}
\end{center}
\vskip0.3truecm
\begin{center}
\leavevmode
\epsfxsize=5.0cm
\epsfbox{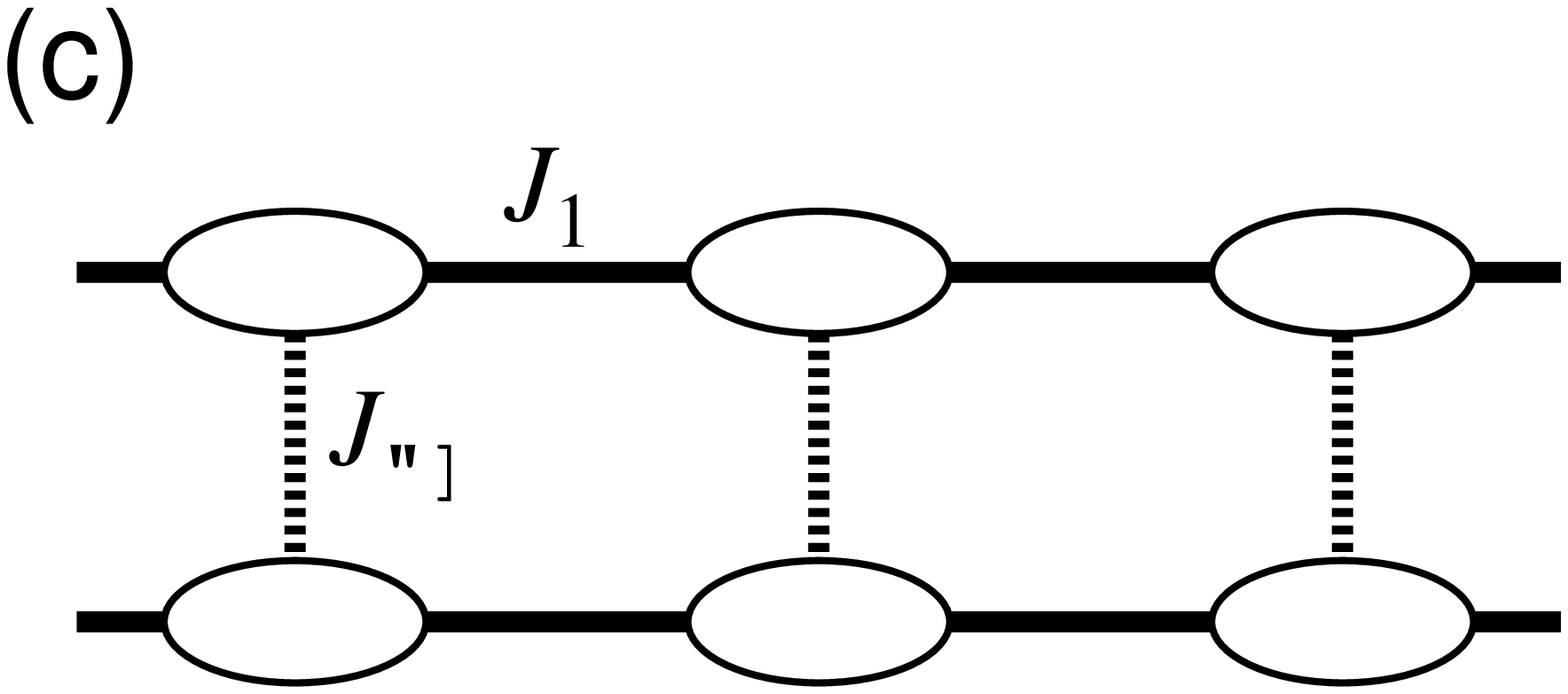}
\label{fig.1}
\end{center}
\caption{(a) Schematic illustration of BIP-TENO. Open circles represent $S=1/2$ spins.
(b) Plane type illustration of BIP-TENO.
(c) Effective model of BIP-TENO in case of $J_{\rm F} \to \infty$.
 Open ellipses represent $S=1$ spins.}
\end{figure}
The plateau at $M=0$ is interpreted as the manifestation of
the Haldane state.
On the other hand, the $M_{\rm s}/4$ plateau is not easily understood.
The necessary condition for the quantized magnetization plateau
by Oshikawa, Yamanaka and Affleck\cite{OYA} is 
\begin{equation}
    Q(S_{\rm unit} - \langle m \rangle) = {\rm integer},
    \label{eq:OYA}
\end{equation}
where $Q$ is the spatial period of the wave function of the state
measured by the unit cell, 
$S_{\rm unit}$ and $\langle m \rangle$ are the total spin
and the magnetization per unit cell,
respectively. 
Applying the theorem to our model, $S_{\rm unit} = 2$ and $m = 1/2$,
this condition is not satisfied if $Q=1$.
Thus, $Q=2$ is required at least,
which means the occurrence of the spontaneous symmetry breaking
in the $M_{\rm s}/4$ plateau state.
The magnetization plateaux caused by the spontaneous symmetry breaking
were discussed for the $M_{\rm s}/2$ plateau of the $S=1/2$ zig-zag 
ladder\cite{Tonegawa1,Totsuka,Tonegawa2}
(equivalent to the bond-alternating chain with next-nearest-neighbor interactions),
the $M_{\rm s}/2$ plateau of the $S=1/2$ ladder,\cite{Mila,Okazaki1,Okazaki2,Okamoto}
and also the $(2/3)M_{\rm s}$ plateau of the $S=1/2$ distorted diamond
chain,\cite{Tonegawa3,Honecker} and so on.

The Hamiltonian of the model of Fig.1(c) is 
\begin{eqnarray}
    \hat H
    &=&  J_1 \sum_{l=1,2} \sum_{j=1}^L \vS_{l,j} \cdot \vS_{l,j+1}
     + J_\perp \sum_{j=1}^L \vS_{1,j} \cdot \vS_{2,j} \nonumber \\
    &&~~~~~ + H \sum_{l=1,2} \sum_{j=1}^L S^z_{l,j}
    \label{eq:Ham}
\end{eqnarray}
where $\vS$ denotes the spin-1 operator, 
$j$ the rung number and $l=1,2$ the leg number.
The last term is the Zeeman energy in the magnetic field $H$.
When $J_\perp=0$, our model is reduced to that of two independent usual $S=1$ 
chains, in which no plateau is expected in the magnetization curve
except at $M=0$. 

Let us consider the opposite limit $J_\perp \gg J_1$
by use of the degenerate perturbation theory.
When $J_1=0$, all the rung spin pairs are mutually independent.
In this case, at $M_{\rm s}/4$, half of the rung spin pairs are
in the state
\begin{equation}
    \psi(0,0) 
    =(1/\sqrt{3})
     (|\uparrow \downarrow \rangle + |\downarrow \uparrow \rangle
      -|00\rangle )
\end{equation}
and the remaining half pairs are in the state
\begin{equation}
    \psi(1,1)
    = (1/\sqrt{2})
      (|\uparrow 0 \rangle + |0 \uparrow \rangle )
\end{equation}
where $\psi(S_{\rm tot},S_{\rm tot}^z)$ is the wave function of
two $S=1$ spins coupled by antiferromagnetic interaction $J_\perp$
with the quantum numbers $S_{\rm tot}$ and $S_{\rm tot}^z$.
These wave functions have lowest energies in the subspace of
$S_{\rm tot}^z = 0$ and $S_{\rm tot}^z = 1$, respectively.
Other possible selections have higher energies. 
The $M_{\rm s}/4$ state is highly degenerate as far as $J_1=0$,
because there is no restriction for the configurations of these two states. 
This degeneracy is lifted up by the introduction of $J_1$.
To investigate the effect of $J_1$, we introduce the pseudo-spin $\vT$ with
$T=1/2$.
The $|\Uparrow\rangle$ and $|\Downarrow\rangle$ states of the $\vT$ spin
correspond to $\psi(1,1)$ and $\psi(0,0)$, respectively.
Neglecting other seven states for the rung spin pairs,
the effective Hamiltonian can be written as
\begin{equation}
    \hat H_{\rm eff}
    = \sum_j \left\{  J_{\rm eff}^{xy} (T_j^x T_{j+1}^x + T_j^y T_{j+1}^y)
                    + J_{\rm eff}^z T_j^z T_{j+1}^z \right\}
    \label{eq:T-Ham}
\end{equation}
in the lowest order of $J_1$, where
\begin{equation}
    J_{\rm eff}^{xy} = (8/3)J_1,~~~~~
    J_{\rm eff}^z = (1/2)J_1
    \label{eq:Del}
\end{equation}
Thus the $M_{\rm s}/4$ plateau problem of the original model
is mapped onto the $M=0$ problem of the usual $T=1/2$ antiferromagnetic
$XXZ$ spin chain with the nearest-neighbor (NN) interaction.
As is well known, 
depending on whether 
$\Delta_{\rm eff} \equiv J_{\rm eff}^z/J_{\rm eff}^{xy} \le 1$
or $\Delta_{\rm eff} >1$, 
the ground state of eq.(\ref{eq:T-Ham}) is either the
spin-fluid state (gapless) or the N\'eel state (gapful),
which correspond to the no-plateau state or the plateau state of the
original model, respectively.

We see $\Delta_{\rm eff} = 3/16$ from eq.(\ref{eq:Del}),
which means no $M_{\rm s}/4$ plateau in the model of Fig.1(c),
at least when $J_\perp \gg J_1$.
As already stated, there is no $M_{\rm s}/4$ plateau also when $J_\perp=0$.
Thus the $M_{\rm s}/4$ plateau will not appear in the whole region of
$J_1/J_\perp$ for the model of Fig.1(c),
although we cannot completely exclude the possibility of the $M_{\rm s}/4$
plateau at the intermediate region.
We note that the absence of the $M_{\rm s}/4$ plateau is also established
by numerical method, as shown later in Fig.4.

\begin{figure}
\begin{center}
\leavevmode
\epsfxsize=5.0cm
\epsfbox{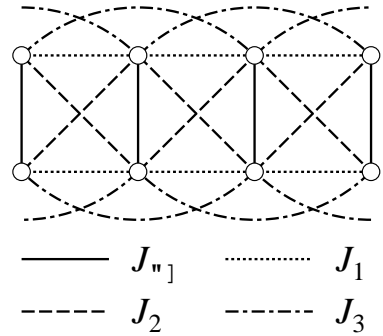}
\end{center}
\label{fig.2}
\caption{$S=1$ ladder model with second and third neighbor interactions.}
\end{figure}

In previous works for the $M_{\rm s}$/2 plateau of the $S=1/2$
spin ladder,\cite{ Mila, Okazaki1, Okazaki2, Okamoto}
it has shown that the second and/or third neighbor interactions bring
about the plateau.
Considering this fact, let us introduce the second ($J_2$) and third ($J_3$)
neighbor interactions to our model, as shown in Fig.2.
Including these interactions,
the effective Hamiltonian (\ref{eq:T-Ham}) is modified to the
generalized version of the $T=1/2$ $XXZ$ Hamiltonian with the NN and the
next-nearest-neighbor (NNN) interactions:
\begin{eqnarray}
    &&\hat H_{\rm eff}
    = \sum_j \left\{  J_{\rm eff}^{xy} (T_j^x T_{j+1}^x + T_j^y T_{j+1}^y)
                    + J_{\rm eff}^z T_j^z T_{j+1}^z \right\} \nonumber \\
    & & + \sum_j \left\{  J_{\rm eff}^{xy(2)} (T_j^x T_{j+2}^x + T_j^y T_{j+2}^y)
                    + J_{\rm eff}^{z(2)} T_j^z T_{j+2}^z \right\} 
      \label{eq:T-Ham2}
\end{eqnarray}
where
\begin{eqnarray}
    &&J_{\rm eff}^{xy} = (8/3)(J_1 - J_2),~~~~~
      J_{\rm eff}^z = (1/2)(J_1 + J_2) \\
    &&J_{\rm eff}^{xy(2)} = (8/3)J_3,~~~~~~~~~~~
      J_{\rm eff}^{z(2)}  = (1/2)J_4
\end{eqnarray}
Note that the $XXZ$ anisotropy parameters of the NN and NNN interactions
are different with each other in general.
In this model, three phases are expected:\cite{NO1,NO2}
these are the spin-fluid state (gapless), the N\'eel state (gapful) and the
dimer state (gapful).

For simplicity, first we consider the $J_2 >0, J_3 =0$ case.
In this case, the NNN interaction in eq.(\ref{eq:T-Ham2}) vanishes
and the anisotropy parameter is 
\begin{equation}
    \Delta_{\rm eff}
    = {3 \over 16}\,{J_1 + J_2 \over J_1 - J_2}
    \label{eq:Del2}
\end{equation}
Therefore the N\'eel condition $\Delta_{\rm eff} >1$ is satisfied when
$J_2/J_1 > 13/19$.
This type of $M_{\rm s}/4$ plateau is due to the N\'eel mechanism 
and called plateau A from now on.

Next we consider the $J_2 = 0, J_3 > 0$ case.
In this case, the anisotropy parameters of the NN and NNN interactions
have the same value $\Delta_{\rm eff} = 3/16$ and
$J_{\rm eff}^{xy(2)}/J_{\rm eff}^{xy} = J_{\rm eff}^{z(2)}/J_{\rm eff}^{z} = J_3/J_1$.
From the phase diagram of $S=1/2$ $XXZ$ chain with the NN and NNN
interactions,\cite{NO1,NO2}
the ground state of (\ref{eq:T-Ham2}) at $\Delta_{\rm eff}=3/16$
is either the spin-fluid state
or the dimer state depending on whether $J_3/J_1 < 0.31$ or $J_3/J_1 > 0.31$.
This type of $M_{\rm s}/4$ plateau, which is brought about by the
dimer mechanism, is called plateau B.

\begin{figure}
\begin{center}
\leavevmode
\epsfxsize=5.0cm
\epsfbox{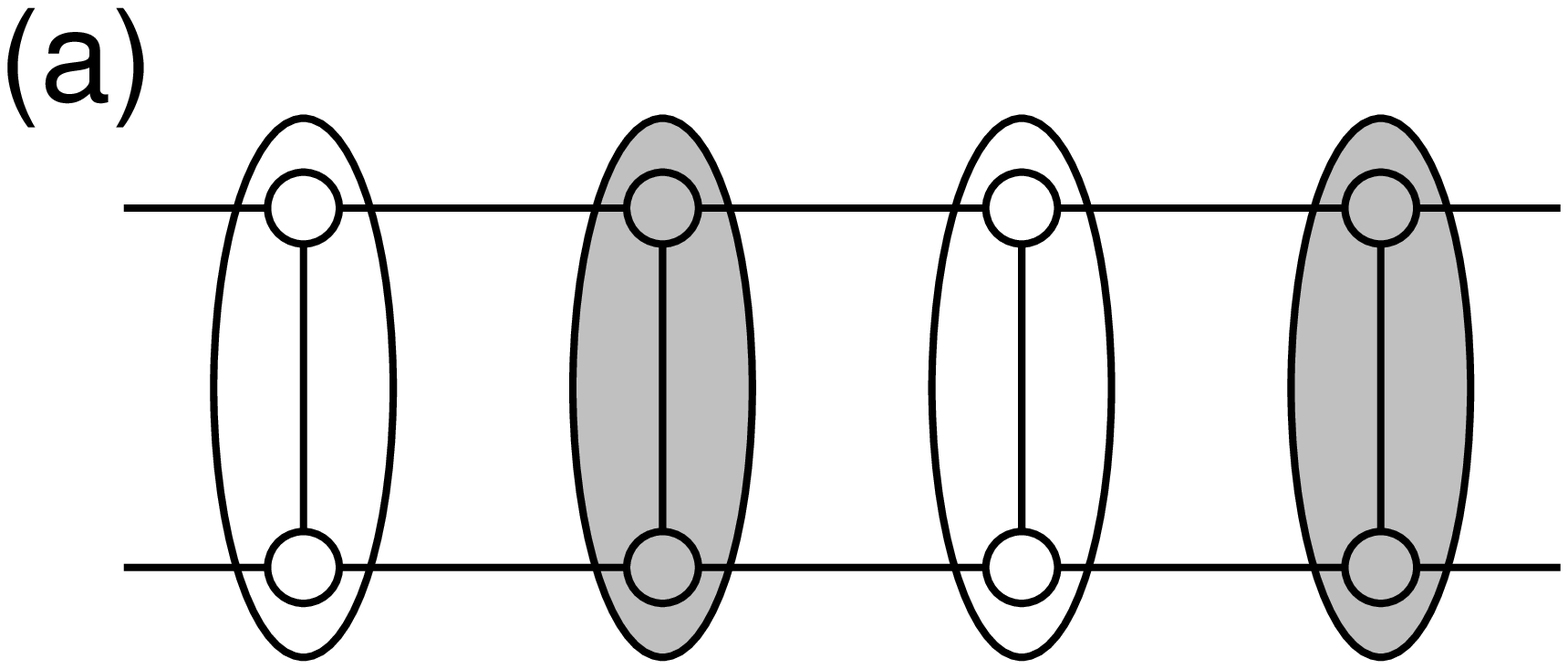}
\end{center}
\vskip0.3truecm
\begin{center}
\leavevmode
\epsfxsize=5.0cm
\epsfbox{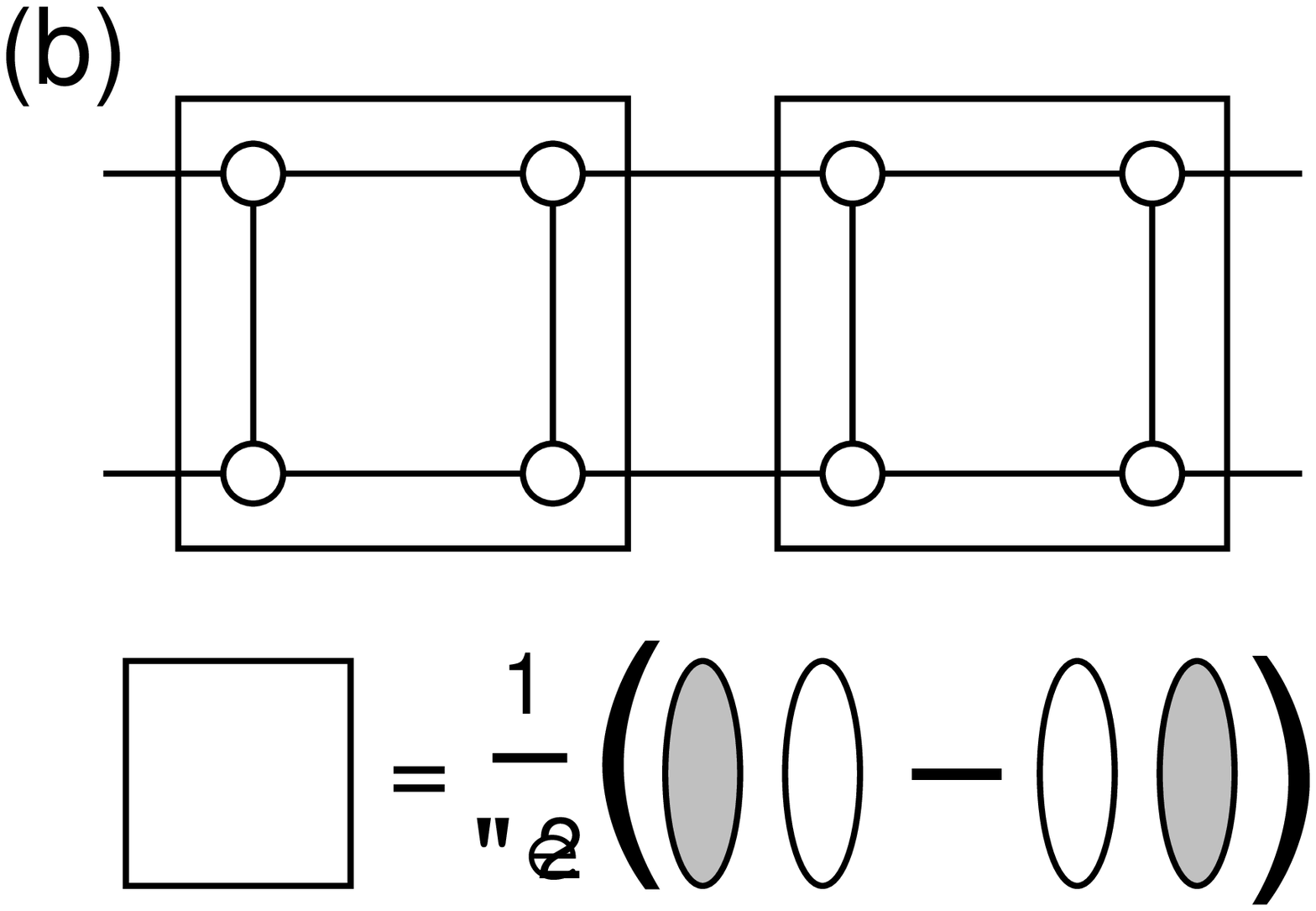}
\end{center}
\label{fig.3}
\caption{(a) Physical picture of the plateau A (N\'eel mechanism).
Open and shaded ellipses represent $\psi(0,0)$ and $\psi(1,1)$
states of the rung pairs ($|\Downarrow\rangle$ and $|\Uparrow\rangle$ in the
$\vT$ picture), respectively.
(b) Physical picture of the plateau B (dimer mechanism).}
\end{figure}
The physical pictures of these two plateau states are shown in Fig.3.
As is clear from Fig.3, both plateau states are doubly degenerate 
due to the spontaneous symmetry breaking.
Then the period of the state is $Q=2$ in both plateau states,
from which we see the realization of the necessary condition eq.(\ref{eq:OYA}).
The transition from the no-plateau state to the plateau A or B state 
(in the language of $\vT$, 
from the spin-fluid state to the N\'eel or dimer state)
is of the Berezinskii-Kosterlitz-Thouless\cite{Berezinskii,KT} (BKT) 
type.\cite{NO1,NO2}
In general case $J_2, J_3>0$,
the transition from the plateau A state to the plateau B state may occur,
which has the Gaussian universality.\cite{NO1,NO2}

We have performed numerical diagonalization by Lanczos method up to 16 spins
for the $S=1$ model with
$J_2$ and/or $J_3$ to confirm to existence of two plateaux,
predicted by use of the degenerate perturbation theory.
As already stated, the transition from the no-plateau state to the
plateau A or B state is of the BKT type, of which boundary is very difficult
to determine from the numerical data when the conventional methods are used.
This is mainly due to the logarithmic size corrections associated with the
BKT transition.
We have used the level spectroscopy
method\cite{Tonegawa2,Okazaki2,Okamoto,Okamoto2,NO2,Nomura},
by use of which the BKT phase boundary can be determined free from
the most dominant logarithmic size corrections.
In the level spectroscopy method, we use three excitations defined by
(for simplicity we are writing the $L=4n$ case, $n$ being an integer)
\begin{eqnarray}
    \Delta_1
    &\equiv& \frac{1}{2}\left\{E_0\left(L,{\frac{L}{2}}+1,\pi\right)
                        +E_0\left(L,{\frac{L}{2}}-1,\pi\right) \right\}
    \nonumber \\
    & &-E_0\left(L,{\frac{L}{2}},0\right), 
    \label{eq:gap1}
    \\
    \Delta_{0\rm A}
    &\equiv& E_{\rm A} \left(L,{\frac{L}{2}},\pi\right)
              -E_0\left(L,{\frac{L}{2}},0\right) 
    \label{eq:gap0A} \\
    \Delta_{0\rm B}
    &\equiv& E_{\rm B} \left(L,{\frac{L}{2}},\pi\right)
              -E_0\left(L,{\frac{L}{2}},0\right)              
    \label{eq:gap0B}
\end{eqnarray}
where $E_0(L,M,k)$ means the lowest eigenvalue of the $S=1$ ladder Hamiltonian 
in the subspace where the eigenvalue of the operator
$S_{\rm tot}^z \equiv \sum_l \sum_j S^z_{l,j}$ is $M$ with the momentum $k$
and system size $L=4n$ under the periodic boundary condition.
$E_0(L,L/2,0)$ is nothing but the ground state energy.
$E_{0\rm A}$ is the lowest energy with $M=L/2,\,k=\pi$ subspace
having the eigenvalue $P = -1$ for the space inversion operation of the
rung number $i \to L - i +1$.
This excitation is called N\'eel excitation.
We have $P=-1$ for $E_{0\rm B}$, which is named dimer excitation.
In the no-plateau state, plateau A state and plateau B state,
the lowest excitations should be $\Delta_1$, $\Delta{0\rm A}$ and $\Delta_{0\rm B}$,
respectively.
Thus the boundaries between these three phases can be determined from the
crossing points among these three excitations
with sweeping the coupling constants.\cite{NO2}

\begin{figure}
\begin{center}
\leavevmode
\epsfxsize=6cm
\epsfbox{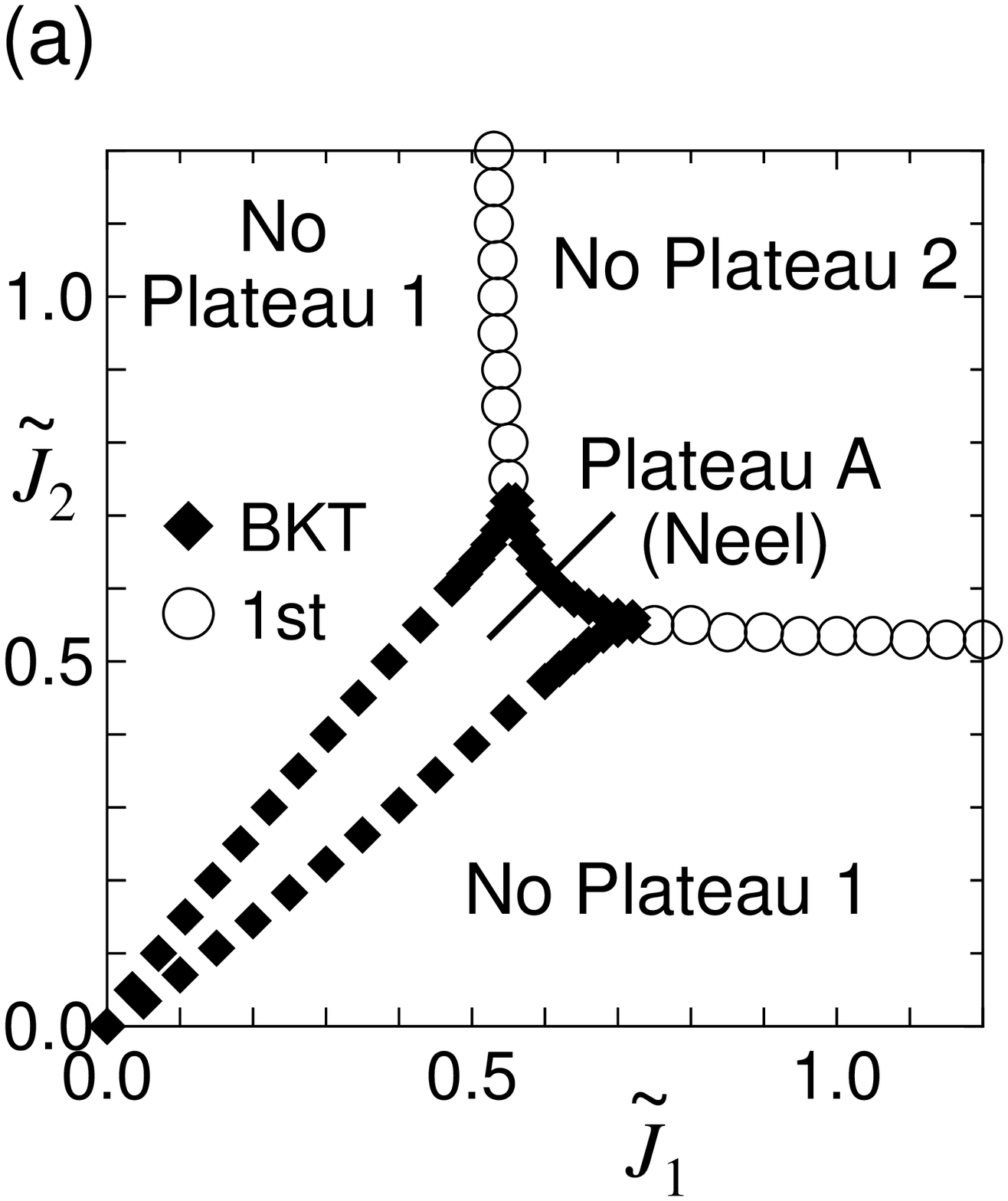}
\end{center}
\begin{center}
\leavevmode
\epsfxsize=6cm
\epsfbox{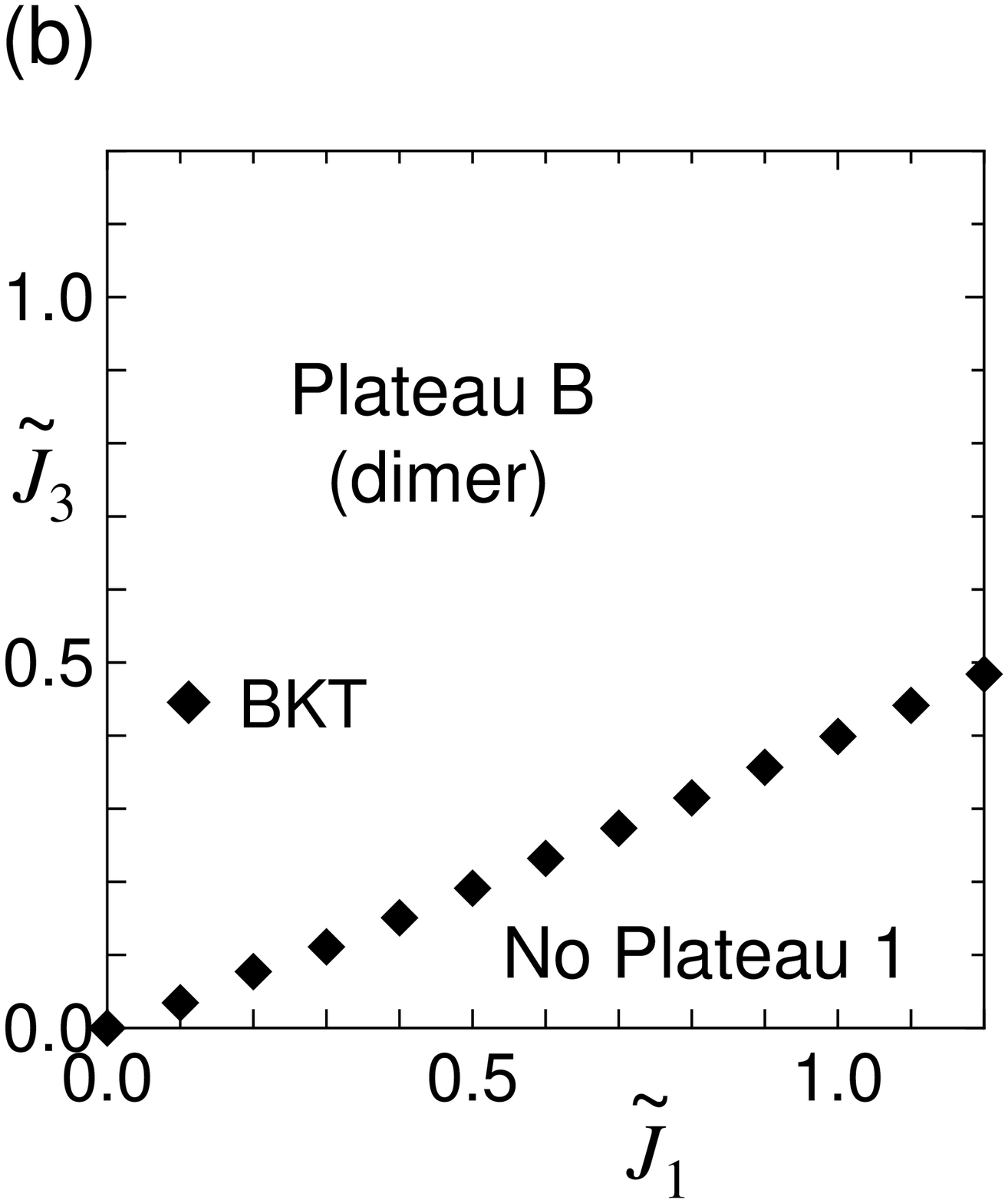}
\end{center}
\label{fig.4}
\caption{The $M_{\rm s}/4$ plateau phase diagram
(a) on the $\tilde J_1- \tilde J_2$ plane when $J_3=0$,
and (b) on the $\tilde J_1- \tilde J_3$ plane when $J_2=0$.
Here $\tilde J_i \equiv J_i/J_\perp$ for $i=1,2,3$.
The closed diamonds represent the BKT phase boundaries and
open circles the first order phase boundaries.}
\end{figure}
\begin{figure}
\begin{center}
\leavevmode
\epsfxsize=5.0cm
\epsfbox{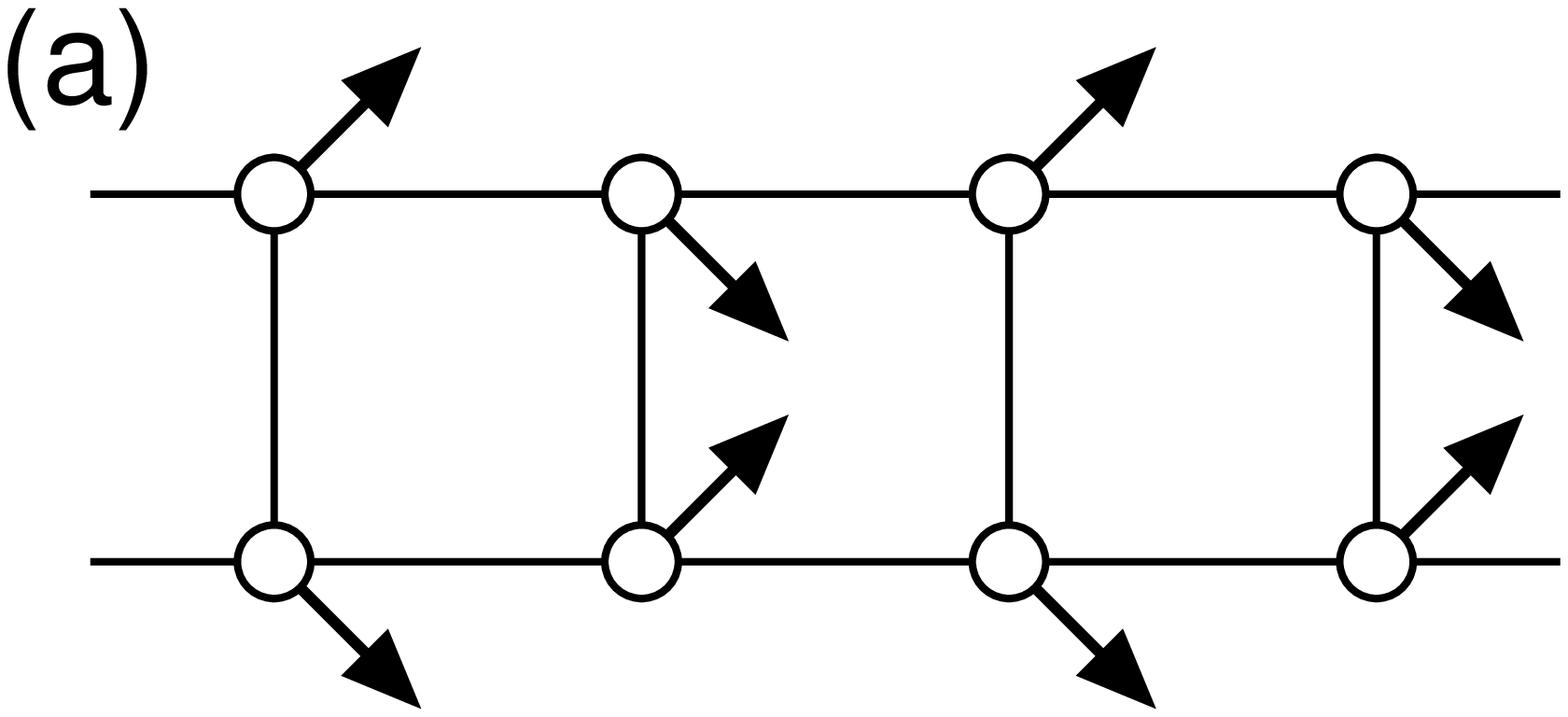}
\end{center}
\vskip0.3truecm
\begin{center}
\leavevmode
\epsfxsize=5.0cm
\epsfbox{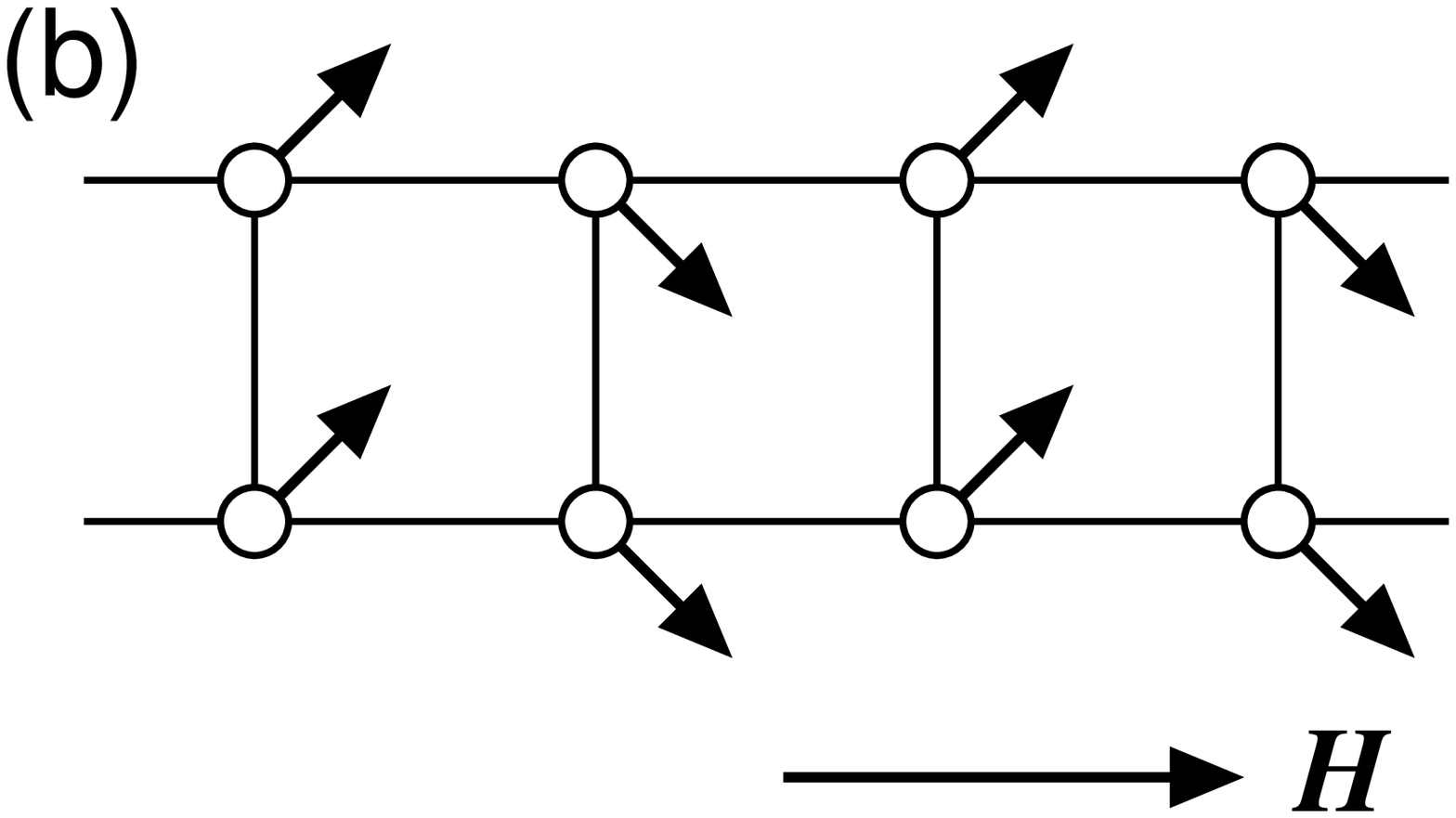}
\end{center}
\label{fig.5}
\caption{(a) Classical picture of (a) the no-plateau 1 state, 
and (b) no plateau 2 state.}
\end{figure}

By use of the level spectroscopy, we obtain the $M_{\rm s}/4$ plateau phase
diagrams in Fig.4.
The estimated error bars of the boundary points are much smaller than
the size of marks.
The line symmetry with respect to the line $\tilde J_2 = \tilde J_1$
in Fig.4(a) is due to the equivalence of the model for interchanging
$J_1 \leftrightarrow J_2$ when $J_3=0$.
The slope of the BKT lines near the origin are very close to 
the predicted values, $13/19$ in Fig.4(a) and $0.31$ in Fig.4(b),
which confirms the high reliability of our analysis of the numerical data
by use of the level spectroscopy method.
In Fig.4(a), there are two no-plateau phases,\cite{Okazaki1,Okazaki2,Sakai}
of which classical pictures are shown in Fig.5.
Note that the long range order is destroyed by quantum fluctuations
in no-plateau states.
The boundaries between two no-plateau states are determined by the
crossing of the ground state.

Following the present study, the critical value of $J_3$ for the realization
of the plateau when $J_2 =0$ is smaller than that of $J_2$ when $J_3=0$,
irrespective of the value of $J_1$.
Furthermore the plateau B phase can appear for any value of $\tilde J_1$,
although the plateau A cannot be realized for larger values of $\tilde J_1$.
Thus the plateau B is easier to be realized than the plateau A.

One may think that there should  exist dashed-line interactions 
in Fig.6(a).
The model of Fig.6(a) has four bond-alternating $S=1/2$ chains.
At a glance, the unit cell seems to consist of eight $S=1/2$ spins,
which may brings about the $M_{\rm s}/4$ plateau without any spontaneous
symmetry breaking.
However, the model of Fig.6(a) can be re-drawn in the plane form of Fig.6(b),
indicating that the unit cell is still composed of four $S=1/2$ spins.
Thus this new interaction cannot be the direct reason for the $M_{\rm s}/4$
plateau.
In fact, in the degenerate perturbation theory used in this letter,
this new interaction only modifies $J_{\rm eff}^{xy}$ and $J_{\rm eff}^z$
in eq.(7) without changing their ratio 
$\Delta_{\rm eff} = J_{\rm eff}^{xy} / J_{\rm eff}^z$.
A similar situation can be found in two-leg bond-alternating ladder
investigated by Cabra and Grynberg.\cite{Cabra}

\begin{figure}
\begin{center}
\leavevmode
\epsfxsize=5.0cm
\epsfbox{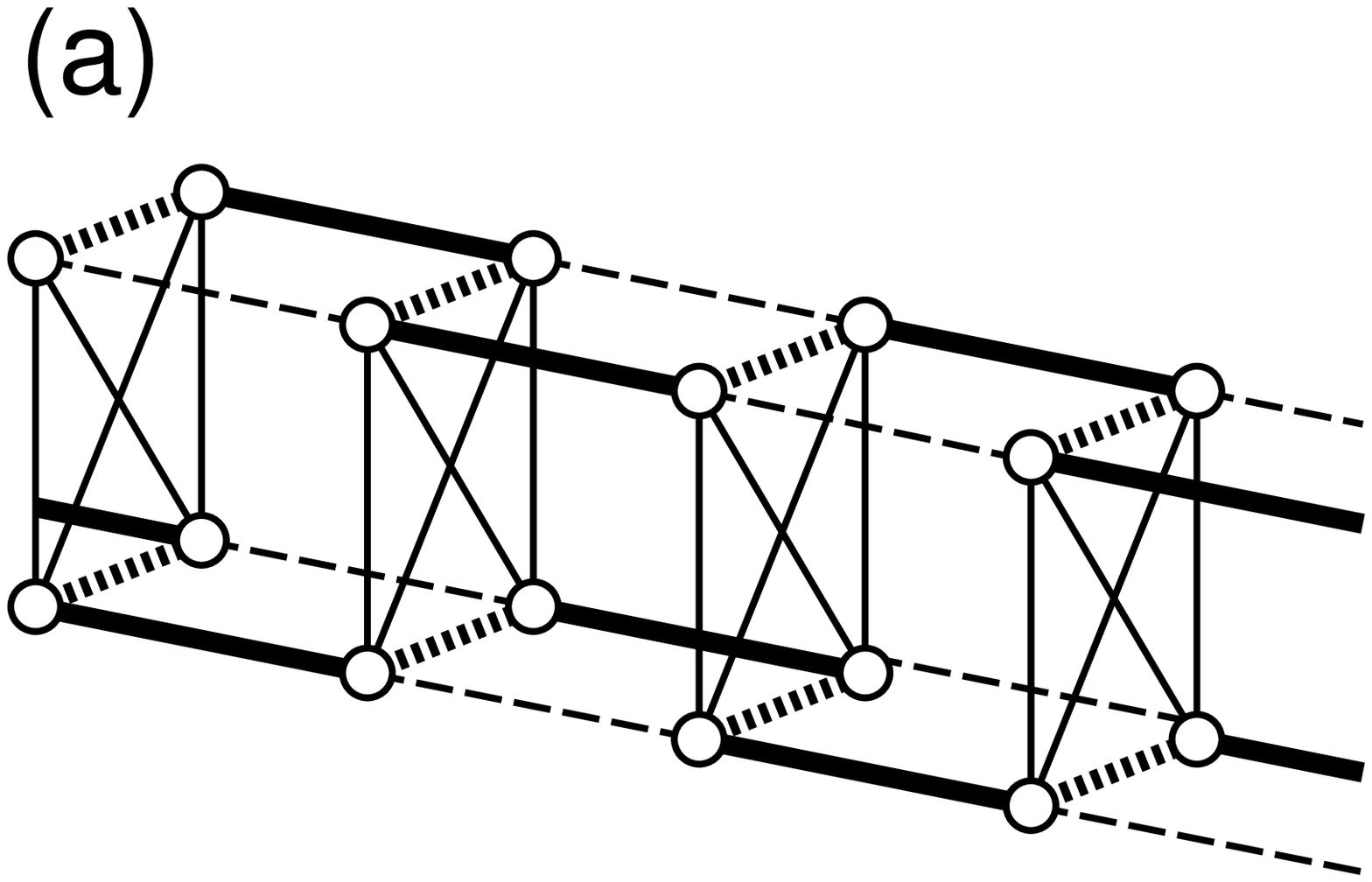}
\end{center}
\vskip0.3truecm
\begin{center}
\leavevmode
\epsfxsize=5.0cm
\epsfbox{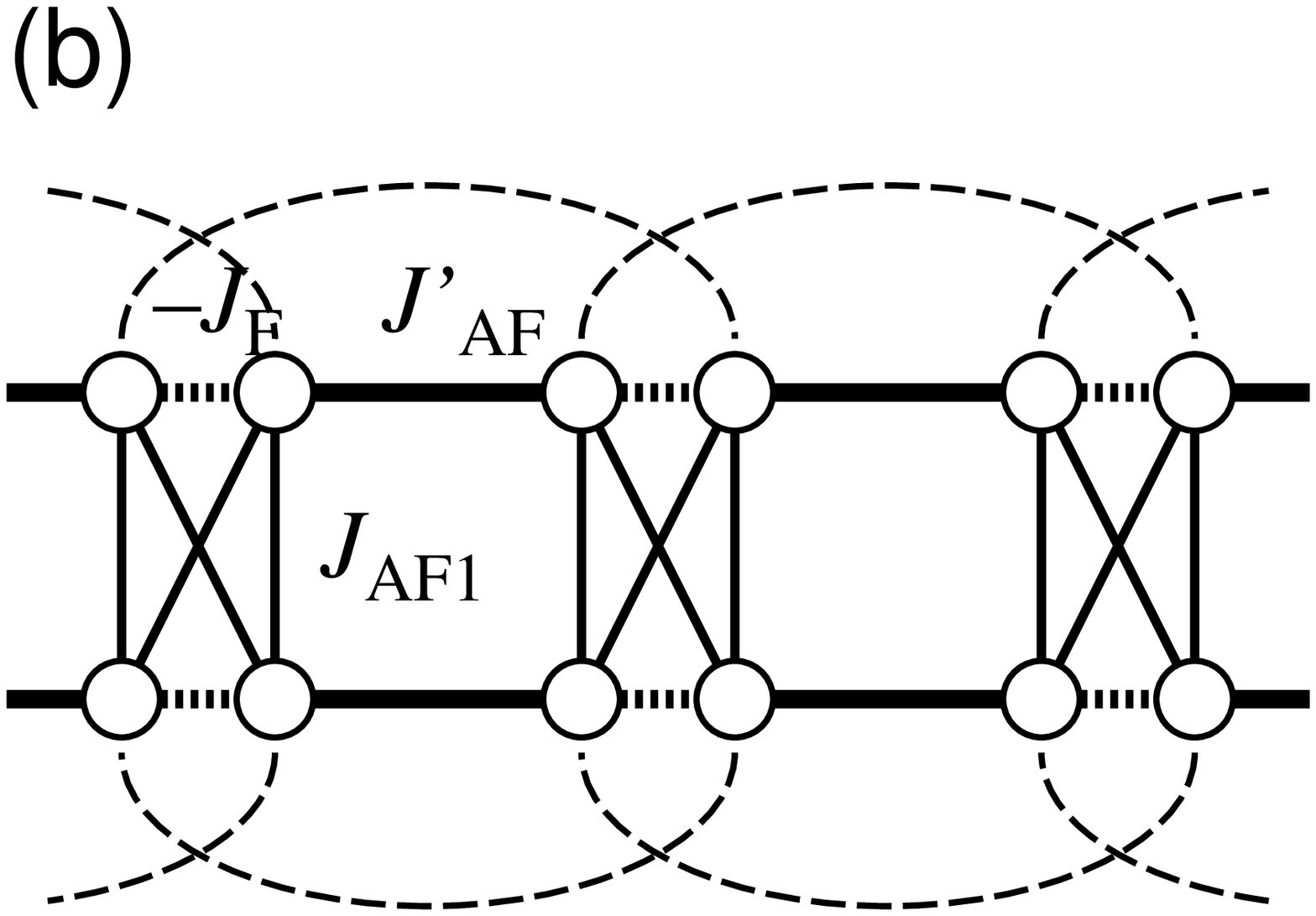}
\end{center}
\label{fig}
\caption{(a) Picture of four bond-alternating $S=1/2$ chains. 
(b) Plane type illustration of (a). The unit cell consists of four $S=1/2$ spins.}
\end{figure}
In conclusion, we have investigated the $M_{\rm s}/4$ plateau of the
$S=1$ spin ladder, proposing two mechanisms for the plateau.
We have also clarified the plateau-no plateau phase diagram by
use of the degenerate perturbation theory, and the level spectroscopy
analysis of the numerical diagonalization data.
The full study of the magnetization process of the present model
will be published elsewhere.

\section*{Acknowledgments}
We would like to express our appreciations to Dr. Yuko Hosokoshi
(Institute for Molecular Science)
and Prof. Tsuneaki Goto (ISSP, University of Tokyo) for stimulating discussions.


\end{document}